\documentclass[a4paper,fleqn,usenatbib,letters]{mnras}

\usepackage{times}

\usepackage[T1]{fontenc}
\usepackage{ae,aecompl}

\usepackage{graphicx}	
\usepackage{amsmath}	
\usepackage{amssymb}	
\usepackage{color}
\usepackage{hyperref}
\title[Radio emission from GX 1+4]{Discovery of radio emission from the symbiotic X-ray binary system GX 1+4}

\author[Van den Eijnden et al.]{
\noindent J. van den Eijnden$^{1}$\thanks{E-mail: a.j.vandeneijnden@uva.nl},
N. Degenaar$^{1}$,
T. D. Russell$^{1}$,
J. C. A. Miller-Jones$^{2}$,
\newauthor R. Wijnands$^{1}$,
J. M. Miller$^{3}$,
A. L. King$^{4}$ and
M. P. Rupen$^{5}$\\
$^{1}$Anton Pannekoek Institute for Astronomy, University of Amsterdam, Science Park 904, 1098 XH Amsterdam, The Netherlands\\
$^{2}$International Centre for Radio Astronomy Research -- Curtin University, GPO Box U1987, Perth, WA 6845, Australia\\
$^{3}$Department of Astronomy, University of Michigan, 500 Church Street, Ann Arbor, MI 48109, USA\\
$^{4}$KIPAC, Stanford University, 452 Lomita Mall, Stanford, CA 94305, USA\\
$^{5}$Herzberg Astronomy and Astrophysics Research Centre, 717 White Lake Road, Penticton, BC, V2A 6J9, Canada
}

\date{Accepted XXX. Received YYY; in original form ZZZ}

\pubyear{2017}

\begin{document}
\label{firstpage}
\pagerange{\pageref{firstpage}--\pageref{lastpage}}
\maketitle

\begin{abstract}
\noindent We report the discovery of radio emission from the accreting X-ray pulsar and symbiotic X-ray binary GX 1+4 with the Karl G. Jansky Very Large Array. This is the first radio detection of such a system, wherein a strongly magnetized neutron star accretes from the stellar wind of an M-type giant companion. We measure a $9$ GHz radio flux density of $105.3 \pm 7.3$ $\mu$Jy, but cannot place meaningful constraints on the spectral index due to a limited frequency range. We consider several emission mechanisms that could be responsible for the observed radio source. We conclude that the observed properties are consistent with shocks in the interaction of the accretion flow with the magnetosphere, a synchrotron-emitting jet, or a propeller-driven outflow. The stellar wind from the companion is unlikely to be the origin of the radio emission. If the detected radio emission originates from a jet, it would show that that strong magnetic fields ($\geq 10^{12}$ G) do not necessarily suppress jet formation.
\end{abstract}




\section{Introduction}

GX 1+4 is an accreting X-ray pulsar with a long, $\sim 120$ second rotation period, which was discovered in 1970 with balloon X-ray experiments \citep{lewin71}. \citet{glass73} identified its optical counterpart as the M6III-type red giant V2116 Oph. The companion star orbits the pulsar most probably in a wide, $1161$ day orbit \citep{hinkle06,ilkiewicz17}, although a shorter orbital period of $\sim 304$ days has also been claimed \citep{cutler86}. The magnetic field strength of GX 1+4 is debated: standard disk accretion theory predicts a magnetic field $B \sim 10^{13}-10^{14}$ G \citep{dotani89,cui04}, while both \citet{rea05} and \citet{ferrigno07} marginally detected a possible cyclotron line implying a field of $B \sim 10^{12}$ G. While the former estimate would be among the largest inferred field strengths in accretion-powered slow pulsars, the latter is more typical for this class of sources.

Jets, strongly collimated outflows, are ubiquitous in accreting systems. Accreting black holes (BHs) show a correlation between their X-ray emission (tracing the accretion flow), their radio emission (tracing the jets), and black hole mass, spanning over eight orders of magnitude in mass \citep{merloni03,falcke04, plotkin13}. The subset of neutron star (NS) low-mass X-ray binaries (LMXBs) that accrete through Roche-lobe overflow of the companion, has been suggested to follow similar correlations, although no universal relation has emerged \citep[][]{migliari06,tudor17,gusinskaia17}. Despite the ubiquity of jets, their formation and collimation is still poorly understood. Hence, comparing jet properties between classes of accreting systems can reveal relevant parameters and necessary conditions for jet formation. For instance, no jets have to date been observed in accreting NS systems where the magnetic field strength exceeds $\sim 10^9$ G \citep{fender97,fender00,migliari06, migliari11,migliari12}. Indeed, theoretical arguments exist in support of the suppression of jet formation by stronger magnetic fields \citep{meier01,massi08}.

A class of systems where (radio) jets have not yet been detected are the symbiotic X-ray binaries (SyXRBs), which are NS LMXBs accreting from the stellar wind of a M-type giant donor. With V2116 Oph as its optical counterpart, GX 1+4 was the first discovered SyXRB, although the larger object class of SyXRBs has only emerged in the past decade \citep[e.g.][]{masetti06,masetti07a,masetti07b,nespoli08,bahramian14b,kuranov15}. Currently, $8$ accreting NSs have been confirmed as SyXBRs, and $6$ more have been proposed \citep[e.g.][]{bahramian14b,kuranov15}. More commonly observed than SyXRBs are their white-dwarf equivalents, called symbiotics or symbiotic stars (SySts), of which around $200$ sources are known \citep[e.g.][]{flores14}. Jets have been observed in several SySts, both through resolved radio imaging, and X-ray or optical spectroscopy \citep[see][for a list]{brocksopp04}.

In this Letter, we report on the discovery of radio emission from the SyXRB GX 1+4 using the Karl G. Jansky Very Large Array (hereafter VLA). This detection constitutes both the first radio detection of a SyXRB and the first hints of a jet from a accreting X-ray pulsar with a strong magnetic field. After describing the observations and the results, we will consider the nature of the radio emission, compare the detection with other classes of accreting systems and discuss the implications for jet formation mechanisms. 
\section{Observations}

\subsection{Radio}

We observed GX 1+4 with the VLA (project code: VLA/13A-352, PI: Degenaar) on 16 June 2013 from 06:24:49 -- 07:08:41 UT for an on-source time of $\sim 27$ min, as part of a larger program studying persistent X-ray bright NS LMXBs. We observed at X-band in the frequency range from $8$ to $10$ GHz. The primary calibrator used was J1331+305, while the nearby phase calibrator was J1751-2524 (angular separation: $4.5^{\rm o}$). During the observation, the VLA was in C-configuration, corresponding to a resolution of a $4.05"\times1.80"$ synthesized beam (with a position angle of $1.59^{\rm o}$). 

We used the Common Astronomy Software Applications package (CASA) v.4.7.2 \citep{mcmullin07} to calibrate and image the data. There was no significant RFI during the observation. We imaged Stokes I and V using the multi-frequency, multi-scale \textsc{clean} task, with Briggs weighting and a robustness of zero. We note that we did not explicitly apply a polarization calibration, which means that our estimates of the circular polarization might be affected by beam squint at the level of a few percent. In the absence of bright radio emission in the field, we did not apply any self-calibration. We reached an RMS noise of $\sim 7.3$ $\mu$Jy\,beam$^{-1}$, corresponding to a $3\sigma$ threshold of $\sim 22$ $\mu$Jy\,beam$^{-1}$ for a detection. As we expect LMXBs to be point sources at the available resolution, we determined flux densities by fitting in the image plane using the \textsc{imfit} task, forcing an elliptical Gaussian with the same FWHM as the synthesized beam. Additionally, we searched for time and frequency variability; we imaged each of the three target scans separately to search for time-variability and imaged the $8$--$9$ and $9$--$10$ GHz bands separately to place constraints on the radio spectrum of GX 1+4. 

\subsection{X-ray}

We searched the archives for X-ray observations of GX 1+4 contemporaneous with the VLA radio observations. Only the \textit{MAXI} monitoring telescope onboard the International Space Station \citep{matsuoka09} performed X-ray observations of GX 1+4 around the time of our radio epoch. We downloaded the $2$--$20$ keV Gas Slit Camera (GSC) spectrum for entire day of 16 June 2013 from the \textit{MAXI} website (\href{http://maxi.riken.jp}{http://maxi.riken.jp}). Despite the relatively low spatial resolution of \textit{MAXI}, no known bright X-ray sources were located in the automatically selected source extraction region so our obtained source spectrum should be free of contamination by any source. The background region does overlap with the ultra-compact X-ray binary SLX 1735-269, which was undetected by \textit{MAXI} during the VLA epoch, and the NS LMXB GX 3+1. The background spectrum might thus be slightly contaminated.

\section{Results}

\subsection{Radio}

\begin{figure}
  \begin{center}
    \includegraphics[width=\columnwidth]{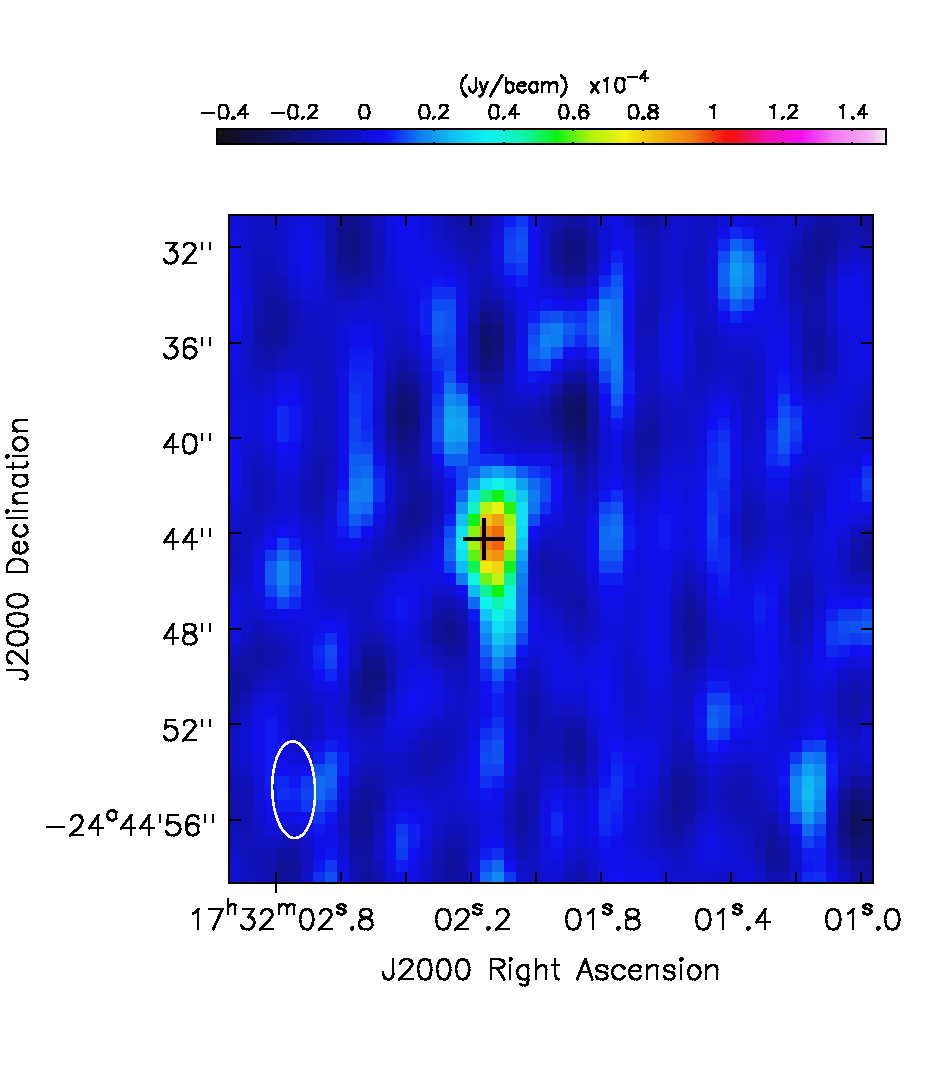}
    \caption{VLA 9 GHz image of GX 1+4. The black cross shows the most accurate position of GX 1+4, from 2MASS (near-infrared), which is accurate to $0.1$ arcsec. The half-power contour of the synthesized beam is shown in the bottom left corner.}
    \label{fig:image}
  \end{center}
\end{figure}

Fig. \ref{fig:image} shows the zoomed, full-bandwidth VLA target image of the entire observation. GX 1+4 is detected at a $\sim 14.4\sigma$ significance with a radio flux density $S_{\nu} = 105.3 \pm 7.3$ $\mu$Jy at $9$ GHz. Assuming a distance of $D=4.3$ kpc \citep{hinkle06} and using $L_R = 4\pi\nu S_{\nu}D^2$, this corresponds to a luminosity of $L_R = (2.10 \pm 0.15)\times10^{28}$ erg s$^{-1}$. GX 1+4 is not detected in Stokes V, implying a $3\sigma$ upper limit of $21\%$ on the circular polarization. This rules out a coherent emission mechanism, as then a $\sim 100\%$ circular polarization is expected. The target is also significantly detected in each of the separate $8$--$9$ GHz and $9$--$10$ GHz bands, at $109 \pm 11$ and $93 \pm 10$ $\mu$Jy respectively. Due to the small frequency range and relatively large uncertainties, the spectral index $\alpha$ (where $S_{\nu} \propto \nu^{\alpha})$ remains poorly constrained at $\alpha = -0.7 \pm 3.3$. We do not detect any significant intra-observational time variability, as the flux densities for each individual source scan ($96 \pm 13$, $82 \pm 13$ and $106 \pm 13$ $\mu$Jy) are consistent within their uncertainties. 

The full observation provides a source position of ${\rm RA} = 17^{\rm h}32^{\rm m}02^{\rm s}.13 \pm 0^{\rm s}.008$ and ${\rm dec} = -24^{\rm o}44'44''.37 \pm 0''.28$. The uncertainties on this position are calculated by dividing the synthesized beam size by the S/N ratio of the detection. Our position is consistent with the best known source position from the near-infrared 2MASS survey \citep[][object name J17320215-2444442]{skrutskie06}, which has a positional accuracy of $0''.1$. We overlay this near-infrared position with the radio image in Fig \ref{fig:image}.

\subsection{X-ray}

We estimate the X-ray flux and luminosity by fitting the \textit{MAXI} GSC spectrum. The GSC spectrum contains $\sim 850$ counts, sufficient for a basic model fit in \textsc{xspec}. We model interstellar absorption using \textsc{tbabs}, assuming abundances from \citet{wilms00}. An absorbed \textsc{powerlaw}-model does not provide a decent fit, with a $\chi^2$ of $34.7$ for $12$ degrees of freedom. An absorbed \textsc{bbodyrad}-model instead returns a better fit with a unabsorbed $0.5$--$10$ keV flux of $(4.6 \pm 0.6)\times10^{-10}$ erg s$^{-1}$ cm$^{-2}$ and a $\chi^2$ of $5.4$ for the same number of parameters. For a distance of $4.3$ kpc \citep{hinkle06}, this flux corresponds to an X-ray luminosity of $L_X = (2.0 \pm 0.2)\times10^{36}$ erg s$^{-1}$, of the order of $\sim 1$ percent of the Eddington Luminosity. While a blackbody model most accurately describes the spectrum, it is difficult to assign an X-ray state to the source as it is unclear whether there is an accretion disk, the spectrum is of low quality and GX 1+4 does not show canonical NS LMXB states.

\section{Discussion}

We report a $105.3 \pm 7.3$ $\mu$Jy ($>14\sigma$ significance) radio detection of the SyXRB GX 1+4 with the VLA at $9$ GHz. Previously, \citet{seaquist93} and \citep{fender97} reported non-detections with upper limits of $90$ $\mu$Jy at $8.3$ GHz and $240$ $\mu$Jy at $2.3$ GHz, respectively. \citet{marti97} presented a marginal $3\sigma$ detection of the source ($60\pm20$ $\mu$Jy at $5$ GHz) with the VLA from the best known optical position of GX 1+4 at the time, but did not claim a significant detection. As their radio position is consistent with ours, it is likely that they indeed observed GX 1+4. Our analysis presents the first significant detection of radio emission from GX 1+4.

\citet{manchanda93} reported the detection of two radio sources approximately equidistant to ($\sim 4$ arcmin) and aligned with the optical position of GX 1+4, proposing that these two emission regions could be radio lobes from a jet. Similar to \citet{marti97} and \citet{fender97}, we detect these two emission regions. However, we measure their positions to be consistent with the \citet{manchanda93} detections, implying they do not move away from the binary system. Furthermore, as already noted by \citet{fender97}, one of the two sources is resolved into two individual sources, and no radio emission is detected connecting them to GX 1+4. Hence, they are likely unrelated to GX 1+4 and their alignment is probably coincidental. 

The origin of the detected radio emission is not immediately obvious. Coherent emission, which would be $\sim 100$\% circularly polarized, is ruled out by the non-detection in Stokes V implying a $21\%$ upper limit on the circular polarization. Here, we will discuss several alternative possibilities: the stellar wind, shocks in the magnetosphere, a jet and a propeller-driven outflow. Since the evidence for the cyclotron line in GX 1+4 is weak at best \citep{rea05, ferrigno07}, we will consider these mechanisms for both the magnetic field implied by the cyclotron line ($\sim 10^{12}$ G, typical for a slow X-ray pulsar), and the maximum field inferred from standard disk theory ($\sim 10^{14}$ G). 

As \citet{fender00} suggested for the marginal radio detection by \citet{marti97}, the radio emission could arise from the stellar wind in GX 1+4. A similar scenario has been postulated for the radio detection of the high-mass X-ray binary GX 301-2, which accretes from the wind of a B-hypergiant with a very high mass loss rate of $\sim 10^{-5}$ $M_{\odot}$ yr$^{-1}$ \citep{pestalozzi09}. For GX 1+4, we can estimate the radio flux density from free-free emission from a stellar wind using \citep[][]{wright75}:
\begin{equation}
\begin{split}
S_{\nu} = 7.26\left(\frac{\nu}{10 \rm GHz}\right)^{0.6} &\left(\frac{T_e}{10^4 \rm K}\right)^{0.1}\left(\frac{\dot{M}}{10^{-6} M_{\odot}/\rm yr}\right)^{4/3}\\
&\left(\frac{\mu_e v_{\infty}}{100 \rm km~s^{-1}}\right)^{-4/3}\left(\frac{d}{\rm kpc}\right)^{-2} \rm mJy
\end{split}
\end{equation}
where $\nu$ is the observing frequency, $T_e$ the electron temperature, $\dot{M}$ the mass-loss rate, $\mu_e$ the mean atomic weight per electron and $v_{\infty}$ the terminal wind velocity. We assume that $\mu_e = 1$ (i.e. a pure hydrogen wind) and set $v_{\infty} = 100$ km s$^{-1}$, consistent with measurements in cool, evolved stars \citep{espey08}, the escape velocity of the donor and the wind velocity estimates in GX 1+4 by \citet{chakrabarty97b} and \citet{hinkle06}. For our basic estimate, we ignore the $T_e$ term due to its low power and hence negligible effect on the radio flux. 

To estimate the mass-loss rate, we use the semi-empirical relation for cool, evolved (for instance M-type) stars given by \citet[][but see also \citealt{espey08}]{reimers87}:
\begin{equation}
\dot{M} = 4\times10^{-13} \eta (L/L_{\odot})(g/g_{\odot})^{-1}(R/R_{\odot})^{-1} M_{\odot}~\rm yr^{-1}
\end{equation}
where $1/3 < \eta < 3$ is a dimensionless scaling to account for the uncertainty in $\dot{M}$ measurements, $L$ is the bolometric luminosity, $g \propto M/R^2$ is the surface gravity, and $R$ and $M$ are the stellar mass and radius. For the range of stellar parameters of GX 1+4 as found by \citet{chakrabarty97b}, we estimate $3\times10^{-9} M_{\odot}$ yr$^{-1}$ $\lesssim \dot{M} \lesssim 7\times10^{-8}$ $M_{\odot}$ yr$^{-1}$. This yields a wind radio flux of $0.14$ $\mu$Jy $\lesssim S_{\nu} \lesssim 11$ $\mu$Jy. Since we detect radio emission at a level of $\sim 100$ $\mu$Jy, the stellar wind is unlikely to account for the detected flux. However, caution should be exercised when estimating the mass-loss rate in late type M stars; while the mass-loss rate is lower than in the B-type hypergiant in GX 301-2, and can be estimated through the equation above, the mass-loss rates for the M-type giant donors in SyXRBs and SySts are poorly known and estimates span orders of magnitude from $\sim 10^{-10}$ to $\sim 10^{-5}$ $M_{\odot}$ yr$^{-1}$ \citep{espey08,enoto14}. 

Alternatively, we might observe radio emission from shocks as the accretion flow interacts with the magnetosphere. The Compton limit on the brightness temperature of $10^{12}$ K sets a minimum size of the emission region of $\sim 7.5\times10^4$ km. The size of the magnetosphere, set by the radius where magnetic and gas pressure are equal, depends on the magnetic field strength and the mass accretion rate, estimated from the X-ray flux \citep[e.g][equation 1]{cackett09}. For our observed flux and standard NS parameters, GX 1+4's highest estimated magnetic field strength of $10^{14}$ G yields a magnetosphere size of $\sim 2.4\times10^5$ km. Hence, such shocks are compatible with the properties of GX 1+4 if the magnetic field is indeed as high as $\sim 10^{14}$ G. However, if the magnetic field is only $10^{12}$ G, the magnetosphere is smaller than the minimum emission region size.

The radio emission could also be synchrotron emission from a collimated jet: the observed combination of $L_X$ and $L_R$ is in agreement with the radio and X-ray luminosities in a large sample of low-magnetic field accreting NSs \citep[e.g. $\lesssim 10^9$ G;][]{migliari06,migliari11b,migliari12,tetarenko16b,tudor17,gusinskaia17}, where the radio originates from such a jet. However, a jet identification will require independent confirmation through new observations, as we discuss below. Interestingly, while slow disk winds have been inferred in a handful of NSs with magnetic field strengths above $\sim 10^9$ G \citep[e.g.][and references therein]{degenaar14}, no radio jets have been observed in such sources \citep{fender97,fender00,migliari06,tudose10,migliari11,migliari12}. 

\citet{massi08} argue that a jet can only form when the magnetic pressure is lower than the gas pressure at the truncation radius of the disk, as otherwise the field lines will not be twisted by the disk rotation. In this scenario, slow X-ray pulsars would indeed not launch a jet due to their strong magnetic field. If a jet is actually present in GX 1+4, or alternatively in Her X-1, a strong-magnetic field intermediate-mass X-ray binary where we also recently discovered radio emission ({\color{blue} Van den Eijnden et al., submitted}), either this argument is incomplete or an alternate jet launching mechanism occurs for strong magnetic fields.

If we observe a jet, the alternative launching mechanism might be a magnetic propeller. Such an outflow has been inferred from X-ray observations in two high magnetic field X-ray pulsars \citep{tsygankov16b}, and has been proposed earlier for GX 1+4 based on its correlated X-ray flux and pulsation behaviour \citep{cui04}. A magnetic propeller can arise when, at the radius where the magnetic pressure equals the gas pressure, the rotational velocity of the field (the pulsar spin $P$) exceeds the Keplerian velocity of the disk, creating a centrifugal barrier. As the gas pressure depends on the mass accretion rate and thus the X-ray luminosity, we can estimate the maximum X-ray luminosity for a propeller to occur as a function of magnetic field strength and pulsar spin \citep[e.g.][]{campana02}:
\begin{equation}
L_{X,\rm max} \approx 4\times10^{37} k^{7/2} B_{12}^2 P^{-7/3} M_{1.4}^{-2/3}R_{10}^5 \text{ } \rm erg~s^{-1}
\end{equation}
where $k$ is a geometry factor, typically assumed to be $0.5$ for disk accretion, $B_{12}$ is the magnetic field strength in units of $10^{12}$ G, $M_{1.4}$ is the pulsar mass in $1.4 M_{\odot}$ and finally $R_{10}$ is the pulsar radius in $10$ km. 

For the highest magnetic field estimate of GX 1+4 (i.e. $\sim 10^{14}$ G) and standard NS parameters, $L_{X, \rm max} \approx 5\times10^{35}$ erg s$^{-1}$, which differs by only a factor of a few from our observed X-ray luminosity. Indeed, our observed X-ray flux is similar to the X-ray flux where \citet{cui04} infer the onset of the propeller regime (i.e. $3\times10^{-10}$ erg s$^{-1}$ cm$^{-2}$). However, this propeller scenario might be difficult to reconsile with the marginal radio detection by \citet{marti97}, where $L_X$ was an order of magnitude higher \citep{galloway00} and a propeller was thus not expected. Furthermore, if the magnetic field is instead $\sim 10^{12}$ G, $L_{X, \rm max}$ drops by four orders of magnitude and a propellor scenario can be exluded. 

The above calculation assumes disk accretion, although in GX 1+4 it is unclear whether the NS accretes directly from stellar wind or there is a wind-fed disk \citep[e.g.][]{lu12}. In SySts, such a disk is only inferred in 4 out of 10 sources known to launch a jet \citep{sokoloski01}. Using $k=1$, as appropiate for spherical accretion, the maximum inferred magnetic field yields $L_{X, \rm max} \approx 6\times10^{36}$ erg s$^{-1}$, implying that GX 1+4 was in the propeller regime. However, we again note that the cyclotron-line estimate of the magnetic field rules out a magnetic propellor, even with $k=1$. 

While GX 1+4 is the first SyXRB where radio emission is detected, radio emission and jets are more common in SySts, i.e. white dwarf equivalents of SyXRBs. Six SySts show a resolved radio jet, while four more systems have a jet as inferred from X-ray or optical spectroscopy \citep[e.g.][and references therein]{brocksopp04}. Contrary to GX 1+4, the magnetic field of the WD accretor in SySts is typically orders of magnitude weaker than that required to suppress jet formation in the argument of \citet{massi08}: Z And has the strongest inferred magnetic field in jet-forming SySts at $\gtrsim 10^5 $ G. 

The point source nature and the luminosity of the radio emission in GX 1+4 are consistent with the radio properties of the detected SySts: while all are resolved by MERLIN, VLA or ATCA (angular extents of $0.1$-$3$", corresponding to hundreds of AU), five out of six sources are significantly closer than GX 1+4 and their angular extent would not be resolved with the VLA in C-configuration at $4.3$ kpc \citep{padin85, dougherty95,ogley02,brocksopp04,karovska10}. The sixth source, HD 149427, is located at a larger distance and was resolved by ATCA \citep{brocksopp03}, but only because its angular extent is exceptionally large in comparison with the other sources. The SySts show a large range of radio luminosities, spanning up to three orders of magnitude ($\sim 10^{27}$--$10^{30}$ erg s$^{-1}$). GX 1+4 falls well within this range, with a radio luminosity similar to the SySts Z And, AG Dra and CH Cyg. 

While the stellar wind is unlikely to be the origin of the radio emission in GX 1+4, shocks, a jet, and a magnetic propeller all cannot be excluded. More observations are required to better understand the radio emission; for instance, a jet nature can be tested by measuring the spectral index from multiple radio bands, searching for linear polarization or extended structure, or potentially detecting a jet break in the broadband spectrum. By observing the source simultaneously at radio and X-ray wavelength around the X-ray luminosity where the propeller is expected to switch on and off, the magnetic propeller explanation can also be tested: above the X-ray luminosity threshold, the radio is expected to be quenched in this scenario. Furthermore, this first radio detection of a SyXRB warrants follow-up radio observations of this class of sources, to establish whether these sources generally are radio emitters or GX 1+4 is an outlier. Finally, also motivated by our recent radio detection of the strongly magnetized X-ray pulsar Her X-1 ({\color{blue}Van den Eijnden et al., submitted}), the hypothesis that strong magnetic field suppress jet formation should be revisited with deep radio observations of a large sample of such X-ray pulsars. 

\section*{Acknowledgements}

JvdE and TDR acknowledge the hospitality of ICRAR Curtin, where part of this research was carried out, and support from the Leids Kerkhoven-Bosscha Fonds. JvdE and ND are supported by a Vidi grant from the Netherlands Organization for Scientific Research (NWO) awarded to ND. TDR acknowledges is supported by a Veni grant from the NWO. JCAM-J is the recipient of an Australian Research Council Future Fellowship (FT140101082). 




\input{output_gx14.bbl}






\bsp	
\label{lastpage}
\end{document}